# Propellantless space exploration

Roman Ya. Kezerashvili[1,2,3]
[1] *New York City College of Technology,*
*The City University of New York, Brooklyn, NY, USA*
[2] *The Graduate School and University Center,*
*The City University of New York, New York, NY, USA*
[3] *Long Island University, Brooklyn, NY, USA*

(Dated: October 6, 2025)

Propellantless propulsion refers to methods of space travel that do not require onboard propellant, instead relying on natural forces or external energy sources. In this talk I'll review different approaches have explored and discuss Pros and Cons for each approach for interstellar space exploration.

*Gravitational assist* uses planetary gravity to change a spacecraft's speed and direction without fuel. It's effective but limited to specific alignments.

*Solar sails* harness radiation pressure from sunlight for continuous, fuel-free acceleration. While effective over time, they require large, reflective materials that degrade in space. Speed can be enhanced by thermal desorption triggered by solar radiation.

*Magnetic sails* generate thrust by interacting with the solar wind through superconducting loops that produce a magnetic field. They provide lower acceleration compared to solar sails, and their performance depends on the available power and the variability of solar wind conditions.

*Electric sails* utilize charged tethers to repel solar wind protons, producing gradual acceleration. Their effectiveness depends on the successful deployment of very long, lightweight conductive wires. They can achieve higher acceleration than solar sails, and their performance is influenced by available power and solar wind conditions.

Lastly, *quantum effects*, such as the Casimir force, offer a speculative but intriguing route to propellantless propulsion based on the vacuum energy of space.

## I. INTRODUCTION

Today the *de facto* chemical propulsion rocket remains the main vehicle for space exploration. However, this propulsion system faces fundamental limitations. First, the necessity to transport fuel on board imposes prohibitive constraints on the mass-to-payload ratio and the overall economic cost. Second, the maximum velocity that a rocket can reach is strictly limited by the Tsiolkovsky rocket equation [1].

Over the past half-century, chemically powered rockets have enabled remarkable advances in space exploration. Nevertheless, rockets powered by combustion are extremely inefficient and require enormous amounts of fuel. As an illustrative example, following Ref. [2], consider the hypothetical exercise of attempting to propel a single proton of mass $m_f = 1.67 \times 10^{-27}$ kg using the entire mass of the observable universe, $m_0 = 10^{53}$ kg, converted to chemical propellant. Treating the system as an infinitely staged rocket with a typical ejection speed of $v_e \approx 4$ km s$^{-1}$, the rocket equation gives the final proton speed as

$$\Delta v = v_e \ln\left(\frac{m_0}{m_f}\right) = 4 \ln\left(\frac{10^{53}}{1.67 \times 10^{-27}}\right) \approx 735 \text{ km s}^{-1} \approx 0.002 c. \tag{1}$$

This extreme scenario shows that even if an astronomically implausible amount of mass were converted into chemical propellant, the resulting velocity would still be only of order $10^{-3} c$. Thus, chemical propulsion is fundamentally incapable of accelerating even a single proton to relativistic speeds in any practically meaningful context.

To explore distances far beyond our solar system, radically different propulsion strategies are required. Alternative methods—including ion engines, nuclear thermal propulsion, fission-fragment, and even thermonuclear fusion—face intrinsic limitations due to their low inherent energy efficiency, quantified as

$$\varepsilon = \frac{\text{energy released}}{mc^2},$$

the fraction of rest-mass energy converted to useful kinetic energy. Even advanced fusion schemes achieve only $\varepsilon \sim 10^{-2}$ [3], while chemical rockets reach merely $\varepsilon \sim 10^{-10}$ [4]. Antimatter annihilation engines, although theoretically far more efficient, face enormous practical challenges due to confinement and reaction requirements [5].

This motivates the question: can we exploit natural astrophysical sources as propulsion mechanisms for space exploration? A number of concepts have been proposed to circumvent the need to carry propellant, including:

- *Gravity assist:* use a planetary gravity to increase a spacecraft's speed and direction without fuel [6].

- *Solar sails:* large, reflective membranes accelerated directly by the Sun's electromagnetic flux [7–9]

- *Magnetic sails:* structures using induced magnetic fields to deflect and extract momentum from the solar wind [10–12].

- *Electric sails:* charged tethers or fields interacting with streaming solar wind ions [13, 14].

- *Quantum vacuum effects:* speculative approaches that exploit phenomena such as the Casimir force to provide propellantless thrust.

In this article, we review the physical principles underlying these propellantless propulsion concepts and evaluate their respective advantages and drawbacks for interstellar travel. The paper is organized as follows. In Sec. II, we introduce the main astrophysical and field-based sources of propellantless propulsion. A gravity-assist maneuver is discussed in Sec. III. Sec. IV is devoted to solar sails and related ideas that exploit electromagnetic radiation for propulsion. Sections V and VI present the underlying physics of the magnetic and electric sail concepts, respectively. Finally, attempts to harness quantum vacuum energy for propulsion are discussed in Sec. VII. Concluding remarks follow in Sec. VIII.

## II. SOURCES FOR PROPELLANTLESS PROPULSION

### A. Gravitational field in the solar system

The gravitational field in the solar system is dominated by the Sun, which contains more than 99% of the total solar system mass. Its gravitational pull governs the motion of planets, moons, asteroids, and comets, following Newton's law of universal gravitation. Planetary and satellite motions can be accurately described by Keplerian orbits in the framework of classical mechanics, with small corrections from mutual perturbations and relativistic effects. The gravitational influence of planets creates regions of stability and resonance, such as the asteroid belt gaps and Lagrange points, which are important for spacecraft trajectory design. The gravitational fields of planets can be used for gravity-assist maneuvers in space exploration, allowing missions to reach the outer planets and beyond while conserving propellant.

### B. Solar radiation

Solar radiation that interacts with a spacecraft consists of two main components: (i) electromagnetic radiation, and (ii) high-energy particles (electrons, protons, and light ions). The Sun, as a typical main-sequence star, generates energy primarily through proton–proton fusion, with only a negligible contribution from the carbon–nitrogen–oxygen (CNO) cycle. The CNO cycle is one of the two principal sets of nuclear fusion reactions by which stars convert hydrogen into helium, the other being the proton–proton chain. A portion of the released fusion energy heats the outer solar layers, giving rise to electromagnetic radiation across the spectrum.

Outside Earth's atmosphere, the solar irradiance is about 1370 W/m$^2$, with roughly 47% in the visible range (380–780 nm), corresponding to photon energies between 1.6 and 3.3 eV. Infrared radiation ($\lambda > 780$ nm) contributes another 46%, and ultraviolet radiation and more shorter wavelength ($\lambda < 380$ nm) about 7% [15].

In addition, solar activity such as flares, solar wind, and coronal mass ejections (CMEs) generate energetic particle fluxes. Solar flares, caused by magnetic instabilities, release bursts of electrons, protons, and light ions with energies up to MeV levels, reaching temperatures of tens of millions of Kelvin and emitting primarily in X-rays [16, 17]. Near solar-maximum, there are about 15 bright whitelight flares per year. The duration of solar flares varies from minutes to about 2 hours [16, 17], while fewer appear during solar minimum. A large solar flare produces a significant flux of protons with energy more than 10 MeV.

The *solar wind* is a plasma outflow of electrons [18] and protons from the hot corona, with typical speeds of 300–800 km/s, proton energies of 0.2–4.2 keV, and densities of about 10 ions/cm$^3$ at 1 AU [19]. Closer to the Sun, densities scale with the inverse square of distance, reaching $\sim 4 \times 10^9$ ions/m$^3$ near 0.05 AU. In contrast, CMEs are sporadic, ejecting magnetized plasma "bubbles" at 50–2000 km/s with proton energies from tens of eV up to tens of MeV [20].

The solar wind is a continuous outflow of plasma from the solar corona into interplanetary space. It is composed primarily of electrons and protons, with smaller fractions of alpha particles and heavier ions. As the plasma expands radially, it carries with it the Sun's magnetic field, forming the interplanetary magnetic field. At 1 AU, the solar wind typically exhibits velocities of 300–800 km/s, particle densities of about 5–10 cm$^{-3}$, and temperatures of $10^5$–$10^6$ K. This outflow shapes the heliosphere and governs interactions with planetary magnetospheres and atmospheres, giving rise to phenomena such as auroras.

The solar wind, moving outward at supersonic speeds and carrying the Sun's magnetic field, interacts with the interstellar medium, which itself flows supersonically and brings its own magnetic field. The region of interaction is called the *heliosheath*, composed of the termination shock, the heliopause, and the region between them.

The solar wind exhibits fluctuations in speed, density, and temperature, with a mean velocity of about 400 $km/s$ [21]. First predicted by Parker in 1958 [22] and confirmed in 1962, it has since been observed from 0.29 $AU$ to beyond 70 $AU$ from the Sun, both in the ecliptic and over the solar poles. Its dynamic pressure, about $1.6 \times 10^{-9}$ Pa, must be accounted for in solar-wind-based propulsion concepts.

In summary, the radiation environment relevant to spacecraft spans electromagnetic radiation to energetic photons and corpuscular radiation from eV to hundreds of MeV.

Thus, the Sun emits both photons and plasma, with fluxes decreasing as the inverse square of the Sun–spacecraft distance. Compared to solar radiation, the solar wind plays the dominant role in shaping the heliosphere and its interaction with the local interstellar medium.

### C. Quantum vacuum effects for propulsion

While interplanetary and interstellar space is filled with electromagnetic radiation and particle plasma, one can also consider quantum vacuum effects. Quantum field theory predicts that the vacuum is not truly empty but is characterized by fluctuations of the underlying fields. Observable phenomena include the Casimir effect [23, 24], where closely spaced plates experience an attractive force, and the dynamical Casimir effect [25–28], where time-varying boundaries can generate real photons. In principle, such photons carry momentum and could provide thrust if emitted directionally. Therefore, if vacuum fluctuations are converted into real photons, the emitted radiation can provide thrust. Several proposals have considered exploiting quantum vacuum effects for propellantless propulsion.

### III. GRAVITY-ASSIST MANEUVER

Contrary to popular belief, indirect ballistic trajectories that involve close approaches to one or more intermediate planets do not necessarily require longer flight durations than direct transfer orbits. In fact, substantial reductions in both flight time and launch energy can be achieved if the gravitational energy available during a midcourse planetary encounter is used efficiently. From the perspective of a passing spacecraft, the intermediate planet behaves like a moving gravitational field relative to the inertial heliocentric coordinate system. As a result, the spacecraft heliocentric energy may either increase or decrease depending on the geometry of the encounter.

This process, known as the gravity-assist maneuver or swing-by, or flyby, is a technique in which a spacecraft deliberately modifies its trajectory and energy through a close encounter with a moving celestial body. By exploiting this effect, missions can reach distant planets with significantly reduced fuel consumption and shorter travel times.

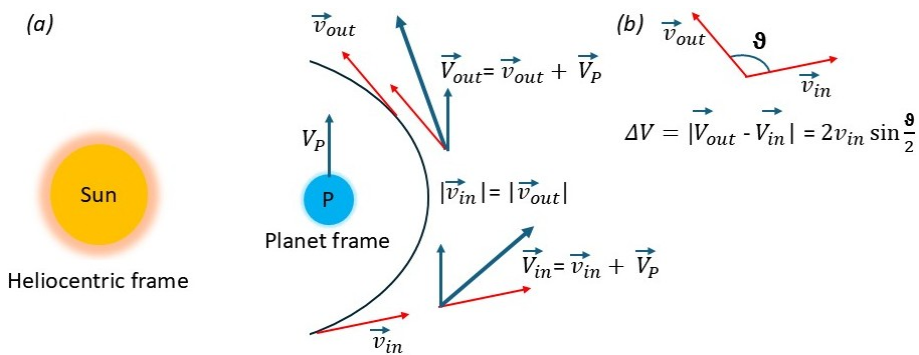

FIG. 1: Illustration of interplanetary mission design in the heliocentric and planet-centered reference frames. The spacecraft velocities in the planet frame are indicated by red vectors, while those in the heliocentric frame are shown by blue vectors. The spacecraft is located within the "sphere of influence" (SOI) of the planet. Inside the SOI, its motion is approximated by a planar Keplerian conic (parabola or hyperbola) relative to the planet.

Among the pioneers of this method, Kondratyuk [6] was one of the first to suggest that spacecraft could "gather velocity" from a moving celestial body. Later, Tsander [8] advanced the theory by deriving quantitative relations and explicitly calculating the maximum energy variations achievable during flybys of different planets. Von Pirquet [29] also employed vector diagrams to describe spacecraft flybys of Jupiter, although he did not explicitly recognize the manoeuvre as a fuel-saving mechanism. Lawden's 1954 paper [30] addresses the optimization of spacecraft trajectories under the influence of gravitational perturbations. The author extends the classical two-body problem by incorporating perturbative forces, such as those arising from third bodies or non-spherical mass distributions, to refine trajectory predictions and minimize fuel consumption. These works laid foundational principles for trajectory design in interplanetary missions, influencing subsequent studies on orbital mechanics and mission planning. However, the phenomenon underlying the swing-by concept has long been known to astronomers [31], and its importance for space exploration has motivated numerous claims of "invention" or pioneering work [32, 33].

Flandro's seminal paper [34], "Fast reconnaissance missions to the outer solar system utilizing energy derived from the gravitational field of Jupiter", laid the foundation for the modern concept of the gravity-assist maneuver in interplanetary mission design. The paper explores how spacecraft can exploit the relative motion and gravitational field of Jupiter (and/or other giant planets) to achieve dramatic increases in heliocentric velocity. This effectively reduces both flight time and propulsion energy requirements for missions to the outer solar system.

In Ref. [35], Prado presents a comparison between two approaches for modeling gravity-assist trajectories: the patched-conics approximation and the restricted three-body problem. The patched-conics method, widely used in mission design, provides a simple analytical framework by dividing the trajectory into conic arcs patched at the sphere of influence. In contrast, the restricted three-body approach accounts for the simultaneous gravitational influence of both the planet and the Sun, yielding a more realistic dynamical description. The study evaluates the differences between the two methods in terms of trajectory geometry, energy change, and accuracy. It is shown that while patched-conics is computationally efficient and adequate for preliminary design, the restricted three-body problem captures dynamical effects neglected in the approximation.

The detailed theory of the gravity-assist manoeuvre and its limitations can be found in standard textbooks [36–42]. In addition, Ref. [33] provides a historical review of the development of gravity-assist theory in the pre-spaceflight era.

The basic idea of the gravity-assist manoeuvre is as follows. In the planet's rest frame, the spacecraft approaches, swings around due to the planet's gravity, and departs with the same speed it had on arrival, i.e.,

$$|\vec{v}_{in}| = |\vec{v}_{out}|,$$

as illustrated in Fig. 1. In this frame, gravity bends the trajectory but does not change the spacecraft's energy.

In the heliocentric reference frame, however, the planet itself moves around the Sun with velocity $\vec{V}_P$. Since the spacecraft effectively "rides along" with this moving gravitational field, it can gain or lose orbital energy relative to the Sun. As shown in Fig. 1, the spacecraft's velocities in the heliocentric frame are given by

$$\vec{V}_{in} = \vec{v}_{in} + \vec{V}_P, \qquad \vec{V}_{out} = \vec{v}_{out} + \vec{V}_P.$$

Inside the sphere of influence (SOI) of the planet, its motion is approximated by a planar Keplerian conic parabola or hyperbola and relative to the planet, the change in velocity of the spacecraft is $\Delta V = |\vec{V}_{out} - \vec{V}_{in}| = 2v_{in} \sin\frac{\theta}{2}$, where $\theta$ is the angle between $\vec{v}_{in}$ and $\vec{v}_{out}$.

The change in specific orbital energy of the spacecraft after a planar flyby is given by [8, 33]:

$$\Delta E = \pm 2v_\infty V_p \sin\left(\delta_{in} \pm \frac{\theta}{2}\right) \sin\left(\frac{\theta}{2}\right), \qquad (2)$$

where $\delta_{in}$ is the angle between the incoming heliocentric velocity and the planet velocity, and $\theta$ is the turning angle of the hyperbolic flyby.

A common question regarding gravity-assist manoeuvres is: from where does the spacecraft obtain the extra energy to increase its heliocentric velocity? The answer lies in the conservation of momentum and energy in the Sun–planet–spacecraft system. The planet transfers part of its orbital momentum around the Sun to the spacecraft. For a prograde encounter, the spacecraft exits the flyby with a higher heliocentric speed and increased orbital energy. The planet, in turn, slows by an immeasurable amount. Thus, the apparent "free" energy gained by the spacecraft originates from the orbital energy of the planet, with gravity acting as the mediator of the exchange.

The gravity assist maneuver was first used in 1959 when the Soviet probe Luna 3 photographed the far side of the Moon. The gravity assist method allows missions to reach distant planets with significantly reduced fuel consumption and shorter flight times. Famous missions such as Pioneer 10 (1972), Pioneer 11 (1973), Voyager 1 and Voyager 2 (1977), Galileo (1989), Ulysses (1990), Cassini (1999), Rosetta (2004), MESSENGER (2004), New Horizons (2006),

Parker Solar Probe (2018), and Solar Orbiter (2020), successfully employed this technique. Voyager 1 launched in 1977 gained the energy to escape the Sun's gravity by performing slingshot maneuvers around Jupiter and Saturn. Voyager 2 trajectory took longer to reach Jupiter and Saturn than its twin spacecraft but enabled further encounters with Uranus and Neptune. Solar Orbiter that launched by ESA in 2020 performed two gravity-assist manoeuvres around Venus and one around Earth to alter the spacecraft's trajectory, guiding it towards the innermost regions of the Solar System.

Table I provides advantages and disadvantages of gravity-assist maneuvers.

TABLE I: Advantages and disadvantages of gravity-assist maneuvers.

| Advantages and Disadvantages | | | |
|---|---|---|---|
| Pros | | Cons | |
| Fuel savings | Enables higher speeds or trajectory changes with minimal propellant. | Launch timing | Requires planetary alignments, limiting launch windows. |
| Velocity boost | Provides large velocity gains via momentum exchange. | Path accuracy | Small errors can cause large deviations in trajectory. |
| Trajectory flexibility | Allows access to distant or otherwise hard-to-reach targets. | Travel time | Indirect paths increase overall mission duration. |
| Proven method | Used in Voyager, Galileo, Cassini, New Horizons, Solar Orbiter. | Navigation | Requires precise calculations and course corrections. |
| Planet science | Flybys enable scientific observations of the assisting planet | | |

To conclude, the gravity-assist manoeuvre transformed space exploration by enabling missions to reach the outer planets and beyond with limited propellant. The spacecraft exits the flyby with a higher heliocentric speed and increased orbital energy. By transferring a small fraction of a planet's orbital momentum to the spacecraft, gravity-assist flybys efficiently reduce flight times and fuel requirements, enabling successful missions to the outer planets and beyond. Its theoretical development reflects a gradual progression: from conceptual insights by Kondratyuk [6] and Tsander [8], to quantitative formalism [30, 34, 35, 43], and eventually to its adoption in mission design by NASA and other space agencies.

## IV. SOLAR SAIL

A solar sail is an elegant propellant-less propulsion system for future exploration of the Solar System and beyond. A solar sail is a large sheet of low areal density material that captures and reflects the Sun's electromagnetic flux as a means of acceleration. The fact that electromagnetic radiation exerts a pressure upon any surface exposed to it was predicted theoretically by James Clerk Maxwell. The main results of the theory of electromagnetic waves were presented to the Royal Society on December 8, 1864, and published in 1865 [44], and later in his famous *A Treatise on Electricity and Magnetism* [45].

According to Maxwell's electromagnetic theory, electromagnetic waves carry energy and linear momentum, and the radiation pressure $P$ exerted on an idealized perfectly reflected flat surface at a distance $r$ from the Sun due to momentum transport by photons is given by

$$P = \frac{2S}{c}, \quad S = \frac{L_\odot}{4\pi r^2}, \qquad (3)$$

where $S$ is the magnitude of the Poynting vector and the solar luminosity is $L_\odot = 3.842 \times 10^{26}$ W.

Pyotr Lebedev first measured light pressure on a solid body in 1899 [46]. He announced the discovery at the 1900 World Physics Congress in Paris, providing the first quantitative confirmation of Maxwell's electromagnetic theory. Nichols and Hull later verified it experimentally, presenting preliminary data in 1901 at the AAAS meeting in Denver and publishing final results in 1903 [47]. Together, these experiments confirmed that light exerts measurable pressure.

46In his 1915 book *Interplanetary Journeys* [7], Yakov Perelman suggested using solar radiation pressure for spacecraft propulsion but deemed it impractical due to its small magnitude. Soviet astronautics pioneer Tsiolkovsky admired Perelman's creativity and wrote the preface to the 1923 edition of the book. The engineering concept of solar sailing was further developed in the early 1920s by Tsiolkovsky and Fridrikh Tsander [9], with Tsander formalizing the theoretical principles in 1924 [8].

The force on a solar sail surface of area $A$ due to the incident solar electromagnetic radiation is given by $F = PA\cos\theta$, where $\theta$ is the angle between the incident solar radiation and the reflective surface normal.

When solar electromagnetic radiation interacts with the material of a solar sail, it undergoes both diffuse and specular reflection, resulting in forces from the momentum exchange of photons. The force due to diffuse reflection acts along the normal to the antisunward surface of the sail, while the force from specular reflection is directed opposite to the reflected radiation. A further contribution arises from absorption of solar radiation, which produces a force along the direction of the incident radiation. In addition, heating of the sail causes part of the absorbed energy to be re-emitted from both the front and back surfaces. This thermal emission generates a net force perpendicular to the sail surface, whose direction depends on the relative emissivities of the two sides.

The total force exerted on a non-perfectly reflecting solar sail due to solar radiation is the result of reflection, absorption, and re-emission by re-radiation. Within the optical model for a non-perfectly reflecting solar sail, following [48, 49], it can be shown that the force due to solar radiation pressure acting on the sail is given by

$$F = \frac{L_\odot}{4\pi c r^2} A \cos\vartheta \left[ a_1 \hat{\vec{r}} - (a_2 + 2a_3 \cos\vartheta)\hat{\vec{n}} \right], \tag{4}$$

with

$$\begin{aligned} a_1 &= 1 - \rho s \\ a_2 &= Bf(1-s)\rho + (1-\rho)\frac{\epsilon_f B_f - \epsilon_b B_b}{\epsilon_f + \epsilon_b} \\ a_3 &= \rho s \end{aligned} \tag{5}$$

In Eqs. (4) and (5) $\hat{\vec{r}}$ is the unit vector in the direction of the heliocentric vector $\vec{r}$, $\hat{\vec{n}}$ is the unit normal vector for the surface area facing the sun, $\vartheta$ is the pitch angle of the solar sail relative to the heliocentric vector $\vec{r}$ and $a_1$, $a_2$, and $a_3$ are dimensionless coefficients that are related to the optical ($a_1, a_3$) and thermo-optical ($a_2$) properties of the solar sail surfaces.

We can conclude that the core physical principle of solar sail propulsion is the transfer of momentum from solar electromagnetic radiation to the sail surface. This process involves: i. Specular and diffuse reflection – photons are redirected, imparting momentum opposite to the reflected direction; ii. Absorption – photons deposit their momentum into the sail, producing a force along the direction of the incident radiation; iii. Thermal re-emission as a secondary effect – absorbed energy is re-emitted, generating an additional force that depends on the emissivity of the sail surfaces. The net effect is the solar radiation pressure force, which acts on the sail and can be controlled by changing the sail's orientation.

Intensive studies on solar sailing since the mid-1970s led to the establishment of the 1st International Symposium on Solar Sailing (ISSS), which took place in June 2007 in Herrsching, Germany. The event focused on recent developments in solar sailing technologies and brought together leading experts in the field. Three years later, in 2010, the 2nd ISSS was held in New York City, USA, where the JAXA team announced the successful deployment of *IKAROS*, the world's first interplanetary solar sail (196 m$^2$), demonstrating spacecraft propulsion using solar radiation pressure. Advances in solar sail concepts, technologies, dynamics and control, and mission design were later documented in a special issue of *Advances in Space Research* (ASR) titled "Solar Sailing: Concepts, Technology, and Missions" [50].

Ten years later, after three additional symposia (2013, 2017, and 2019, held in the UK, Japan, and Germany, respectively) and numerous publications in the field, a second ASR special issue reflected the vibrant state of solar sailing research [51]. In parallel, review papers have been regularly produced on topics related to mission design and applications, attitude control and mission analysis, hardware development and testing, as well as materials and technologies [52–56].

Interestingly, solar sails allow non-Keplerian orbits, enabling persistent coverage of regions that conventional satellites cannot continuously observe. For example, Refs. [57, 58] propose novel solar-sail mission concepts for continuous observation of Earth's polar regions and the lunar surface. Other studies explore using solar sails to maintain Mars-synchronous displaced orbits for remote sensing [59], and placing a solar sail spacecraft into a high-inclination orbit to observe the Sun and heliosphere from out-of-ecliptic angles, providing unique observational vantage points that conventional missions cannot achieve without propellant-intensive maneuvers [60].

Tracking interplanetary or interstellar spacecraft is an important technique for testing the laws of physics. Analysis of spacecraft ephemerides for missions where anomalies have been detected—or could be revealed in the future—may provide data on unknown gravitational and non-gravitational effects. In Ref. [62, 63], it is demonstrated that solar sailcraft can be used to test fundamental physics. In particular, the influence of a special relativistic effect, the Poynting–Robertson effect [64, 65], on various types of solar sail trajectories was considered [66, 67].

It is also of particular interest to analyze how the trajectory of a solar-sail propelled satellite deviates from a geodesic under the action of solar radiation pressure (SRP). Studies of general relativistic effects on bound solar sail orbits, including spacetime curvature and SRP, show deviations from Kepler's third law [68]. Additional deviations arise from frame dragging, the Sun's gravitational multipole moments, a net solar electric charge, and a positive cosmological constant. SRP can amplify these deviations by several orders of magnitude, potentially rendering them detectable. It has been shown how SRP modifies the perihelion shift of non-circular orbits and affects the Lense–Thirring effect [69], which involves precession of polar orbits. Furthermore, the pitch angle for non-Keplerian orbits is influenced by partial absorption of light, general relativistic effects, and the Sun's oblateness. Ref. [70] predicts an analog of the Lense–Thirring effect for non-Keplerian orbits, in which the orbital plane precesses around the Sun.

Finally, consideration of solar-sail propelled satellites within the framework of general relativity—particularly the deflection of escape trajectories due to spacetime curvature and frame dragging near the Sun—leads to significant deviations from Newtonian predictions [71]. Careful analysis of solar sail ephemerides thus provides a method for probing and testing new physics.

Table II gives a balanced list of advantages and disadvantages of solar sail propulsion.

TABLE II: Advantages and disadvantages of solar sail propulsion.

| Advantages and Disadvantages | | | |
|---|---|---|---|
| Pros | | Cons | |
| No propellant | Sunlight provides continuous thrust, enabling long missions. | Low thrust | Acceleration is small, requiring long mission times. |
| Continuous acceleration | Provides velocity gains via persistent force, enabling gradual acceleration. | Large sails area | Thousands of square meters needed to generate useful thrust |
| Scalable system | Larger sails capture more photon momentum, tuning performance. | Deployment challenge | Large sails create structural and deployment challenges |
| Proven method | IKAROS, NanoSail-D2, LightSail-1/2, GAMA Alpha missions. | Material limits | Sails require ultra-thin, reflective, and resistant to space degradation materials |
| Potential for new physics test | Test GR effects such as frame dragging and non-Keplerian orbits scientific observations of planets. | Attitude control | Precise orientation needed for maneuvering |
| Potential for advanced concepts | Driven by sunlight or microwave beams due to thermal desorption | Large sails | Require very thin, fragile materials, hard to deploy. |
| Deep-space capable | Suitable for interplanetary and interstellar missions. | Low thrust | Acceleration is small, requiring long mission times. |

### A. Drag sail

A drag sail is a thin, lightweight membrane deployed to increase the effective cross-sectional area of a spacecraft, thereby enhancing drag forces exerted by the surrounding medium. There are two key concepts for drag sails: (i) aerodynamic drag sails used in low Earth orbit (LEO) for de-orbiting satellites, and (ii) plasma or photon drag sails considered for interplanetary or interstellar missions.

8The underlying physics of drag sail operation relies on three fundamental principles: i) momentum transfer; ii) cross-sectional amplification; iii) energy dissipation. By transferring momentum to ambient atmospheric molecules, plasma ions, or photons, the sail experiences a reduction in velocity. The large deployed surface area of the sail significantly increases the effective interaction area relative to the spacecraft mass. Finally, drag dissipates the spacecraft's orbital or translational kinetic energy, enabling controlled de-orbiting or deceleration.

In general, the drag force in a medium of particles density $n$, particle mass $m_p$, and spacecraft relative velocity $v$ can be expressed as

$$F_d = n m_p v^2 A_{\text{eff}}, \tag{6}$$

where $A_{\text{eff}}$ is the effective interaction area, which may exceed the geometric area if electromagnetic fields extend the sail's reach. For atmospheric drag sail one can write (6) as $F_d = \frac{1}{2} C_d \rho v^2 A_{\text{eff}}$, where $C_d$ is the drag coefficient, $\rho$ is the density of the ambient atmosphere. This equation quantitatively links the sail's surface area and environmental parameters to the resulting deceleration and energy dissipation.

The concept of increasing cross-sectional area to increase drag is very old, used for drag parachutes, and even aero braking in reentry, but those operate in denser atmosphere, not the LEO very thin upper atmosphere/vacuum drag sail context. Therefore, drag sails provide a simple yet powerful mechanism for passive orbital decay in Earth orbit and offer speculative routes for deceleration in interplanetary or even interstellar missions.

The development status of membrane drag-sail deorbit technology for LEO satellites is reviewed, and the current challenges associated with drag-sail devices—such as attitude instability, exposure to the harsh space environment, and folding/unfolding mechanisms—are summarized. To address these challenges, several key technologies have been proposed. These solutions, which provide valuable references for the advancement of drag-enhancing deorbit systems, are introduced and discussed in Ref. [72].

Many countries are actively developing drag-sail technology and conducting in-orbit demonstrations. Research and development of drag-sail (de-orbit sail/drag-augmentation) technology has become a multi-national effort involving universities, national space agencies, and commercial vendors. Notable active programs include efforts in the United Kingdom (Surrey Space Centre, SSTL, RemoveDEBRIS), the United States (NASA and commercial providers), China (state research institutes and launch providers), Japan (JAXA and industry partners), and various European countries coordinated through ESA contracts and national teams. Several of these programs have also conducted in-orbit demonstrations that successfully accelerated the orbital decay of small satellites or upper stages [73–77].

In July 2015, the Surrey Space Center in the United Kingdom launched the DeorbitSail satellite [61], while the University of Glasgow developed the Aerodynamic End of Life DeOrbit System (AEOLDOS) for CubeSats [78]. In 2017, the Surrey Space Centre at the University of Surrey also developed InflateSail as a technical verification satellite [79]. Around the same time, the University of Toronto in Canada arranged four identical triangular drag sails on the CanX-7 satellite, each consisting of a flexible membrane supported by a pair of stiff booms deployed from a compact module [80]. In December 2018, Poland launched the PW-Sat2 satellite into a Sun-synchronous orbit at an altitude of 590 km [81]. In China the first drag sail device of Tianyi Research Institute was launched into orbit [82], followed in September 2019 by the Shanghai Institute of the Shanghai Academy of Spaceflight Technology, which launched the Taurus drag sail test [83]. Japan's Nihon University designed an inflatable membrane sail for active deorbiting [84].

Table III lists representative demonstrations and development programs. NanoSail-D2 (USA) and InflateSail (UK) are among the earliest widely publicized LEO drag-sail demonstrations; RemoveDEBRIS (UK/Europe) included an explicit dragsail de-orbitlet test. China has publicly reported using drag-sail hardware on selected rocket bodies and debris objects to accelerate orbital decay. Many other nations and commercial vendors now offer or evaluate drag-sail modules for CubeSats and small satellites [73–77].

Beyond the Earth's atmosphere, drag sails in interplanetary and interstellar space can be designed to interact with plasma and electromagnetic radiation fields. A major challenge for such missions is achieving sufficient deceleration at the destination, as the spacecraft can reach extremely high speeds during the acceleration phase.

For solar photon sails approaching a target star, the sail can be reoriented to reflect the star's light. A reflective sail oriented opposite to the direction of motion interacts with the stellar photon flux, thereby decelerating the spacecraft [85]. Magnetic drag sails employ large current-carrying loops that generate magnetic fields to deflect charged solar (star) wind particles, producing a drag force [11, 12]. Electric drag sails use long charged tethers that interact electrostatically with solar (star) wind ions, enhancing drag through Coulomb scattering [13, 14].

### B. Beam-driven sail

While solar sails were originally conceived with solar radiation in mind, the invention of lasers in the 1960s enabled the development of efficient laser-sailing propulsion systems. In other words, the concept of using solar electromagnetic radiation for propulsion led to the idea of driving a spacecraft with Earth- or space-based beams reflected from a deployable membrane.

TABLE III: Selected countries, organizations, and in-orbit drag-sail demonstrations. Data are representative, not exhaustive.

| Country | Organization/Operator | Mission/Demonstration | Year |
| --- | --- | --- | --- |
| US | NASA / university teams | NanoSail-D2 (successful sail deployment; accelerated decay) | 2010-2011 |
| UK/Europe | Surrey Space Centre, SSTL; RemoveDEBRIS consortium | InflateSail (Surrey) and RemoveDEBRIS dragsail demo | 2017; 2018–2019 |
| China | SAST and other launch/space organizations | Rocket-body/adapter drag-sail deployments reported for accelerated decay | 2021–2022 |
| Japan | Japan JAXA, industry partners | National and commercial small-sat de-orbit sail development and demonstration plans | mid-2020s |
| EU/ESA | Various companies under ESA | ADEO / industry de-orbit sail contracts and technology demonstrations | 2018–2022 |
| Multi-country | Smallsat vendors | Commercial plug-and-play drag-sail modules for CubeSats (multiple providers) | 2015–present |

Robert Forward was the first to propose a laser-driven sail in 1962 [86], recognizing that a powerful, space-based laser could provide the continuous thrust needed for interstellar missions by transferring momentum to a sail—a concept that requires extremely large lasers to be a viable option. Since then, this idea has been further developed [5, 87–92]. Theoretical analyses have examined the required laser power, sail size, reflectivity, and beam divergence needed for interstellar travel. While technically challenging, such schemes could make interstellar missions feasible in principle under certain optimistic assumptions.

Laser-driven sails have the same physical principle of operation as solar sails, except that solar photons are replaced by "synthetic" ones, e.g., from an Earth-or space-based laser. A collimated light beam impinges on the sail, providing thrust. Significantly higher photon pressures can be achieved than for solar sails, determined by the laser source. The drawback is large mechanical and thermal stresses on the sail membrane, but the benefit is high achievable accelerations for trips to interplanetary targets, and even to interstellar distances within the duration of a human lifetime.

Beam-driven sails have been proposed as candidates for interplanetary and interstellar travel, since they can reach high, even sub-relativistic velocities by reflecting focused, high-powered lasers [2, 5, 93–95] or collimated microwaves [96–101]. Interesting to mention that authors of [97] suggested using thermal desorption by heating the coated sail with microwaves. In [101], it was proposed to employ solar sails coated with materials that undergo thermal desorption at specific heliocentric distances due to solar radiation heating. As the temperature rises, the coating undergoes desorption near the perihelion of the heliocentric escape orbit. This process provides a secondary thrust, boosting the solar sail to escape velocities exceeding 100 km/s [102]. Laser-driven and microwave-driven sails are still at the concept and early experiment stage. No orbital flight tests have yet been made.

For designs targeting significant fractions of the speed of light, such as $0.02c$, relativistic momentum transfer and Doppler shifting of the reflected light must be taken into account. The momentum transfer rate for photons reflecting from a sail moving away at velocity $v$ is reduced by the Doppler factor. In the relativistic regime, the energy required to accelerate the spacecraft to a speed of $0.2c$ can be calculated by applying the conservation of relativistic energy and momentum [2, 5, 95, 103, 104], which is essential for accurate modeling of high-velocity light sail propulsion.

To make the project a reality, several key practical challenges must be overcome. Extremely high continuous or pulsed power is required, and building and operating such arrays, whether ground- or space-based, represents a major engineering and cost challenge. Maintaining the beam on the sail over long distances requires exquisite pointing accuracy and, for laser arrays, precise phasing of segmented transmitters. Achieving ultra-low areal density while ensuring high reflectivity and survivability under intense laser irradiation is difficult. Absorbed power can rapidly heat and damage the sail, which must tolerate this energy without structural or optical failure. Minimizing absorption while maintaining high emissivity is also necessary to prevent overheating. The challenges of relativistic light sails and possible solutions are discussed in Ref. [95].

## V. MAGNETIC SAILS

The magnetic sail is a propellantless propulsion concept that exploits the interaction between a large-scale artificial magnetic field, generated by the spacecraft, and the plasma of the solar wind or interstellar medium. Its fundamental physical principle is the conservation of momentum: incident charged particles, – primarily protons, electrons, and light ions, – are deflected by the magnetic field produced by a device onboard the spacecraft. The deflection of charged particles alters the momentum of the particles and imparts a corresponding reaction force to the sail.

The concept was first introduced by Andrews and Zubrin [10, 11]. The proposed original *MagSail* consists of a large superconducting loop (tens of kilometers in radius) carrying a current $I$, which generates a magnetic dipole moment.

The Lorentz force acting on a solar wind particle of charge $q$, mass $m$, velocity $\mathbf{v}$ in a magnetic field $\mathbf{B}$ is

$$\mathbf{F} = q\,\mathbf{v} \times \mathbf{B}. \tag{7}$$

For a circular loop of radius $R$ carrying current $I$, the on-axis magnetic field at the center is

$$B = \frac{\mu_0 I}{2R}, \tag{8}$$

where $\mu_0$ is the permeability of free space. The effective "magnetopause" radius $R_m$, which defines the region where the magnetic pressure balances the dynamic pressure of the solar wind, can be estimated from

$$\frac{B^2(R_m)}{2\mu_0} \approx \frac{1}{2}\rho v^2, \tag{9}$$

where $\rho$ and $v$ are the mass density and velocity of the solar wind.

The thrust of the magnetic sail can then be expressed as

$$F \sim \pi R_m^2 \, \rho v^2, \tag{10}$$

which corresponds to the momentum flux of the deflected plasma intercepted by the effective cross section.

In Ref. [12], Zubrin extends the concept of the magnetic sail to near-Earth applications, investigating its potential use for escaping from LEO. The underlying physical principle is that a superconducting loop can interact not only with the solar wind but also with Earth's magnetosphere and trapped plasma, producing thrust by deflecting charged particles and currents. The paper derives thrust estimates and evaluates efficiency, showing that while the achievable thrust in near-Earth space is small, continuous operation could gradually raise a spacecraft's orbit and eventually enable Earth escape. The study highlights the advantages of propellantless propulsion but also notes the practical limitations and the technical challenges of deploying and maintaining large superconducting coils in orbit. In Refs. [105, 106], a novel concept was presented for deploying a circular superconducting wire attached to a thin circular membrane. Using classical electrodynamics and elasticity theory, it was shown that the superconducting current in the wire can effectively deploy the wire into a large circular shape.

Subsequent developments of magnetic sails have led to variants such as the Mini-Magnetospheric Plasma Propulsion (M2P2) [107] and the Magnetoplasma Sail (MPS) [108–110], both of which reduce the scale of the current loops by injecting plasma from the spacecraft itself. While the concept of a rotating magnetic field plasma sail was introduced in Ref. [111].

In more general consideration, magnetic sails are a class of propellantless spacecraft propulsion systems that utilize the interaction between the solar wind plasma and artificially generated magnetic fields. Several configurations have been developed since the original concept was proposed, each with different approaches to magnetosphere generation and thrust production. Below we summarize the main types.

Proposed by Winglee [107], the *M2P2* aims to reduce coil size by inflating the magnetic field with injected plasma, creating a large artificial magnetosphere:

$$\beta = \frac{nk_B T}{B^2/2\mu_0} \gg 1, \tag{11}$$

where $\beta$ is the plasma-to-magnetic pressure ratio. In Eq. (11) $n$ is particle number density, $k_B$ is Boltzmann constant, and $T$ is plasma temperature. The inflated magnetosphere deflects the solar wind, allowing thrust production without deploying large structural loops. Although promising in principle, kinetic simulations have shown limitations in thrust scaling.

Developed primarily by Funaki and collaborators [108–110], the *MPS* uses plasma injection to expand a magnetic field generated by a superconducting coil. Compared with M2P2, the MPS is designed with larger coils (10–100 m) to improve coupling between solar wind ions and the artificial magnetosphere, especially under conditions where the ion Larmor radius $r_{Li}$ is small compared to magnetospheric size $L_m$:

$$r_{Li} = \frac{m_i v_\perp}{qB} \ll L_m. \tag{12}$$

The thrust is derived from bow-shock formation and solar wind deflection:

$$F \sim \frac{1}{2} C_D n m_i u_{\text{sw}}^2 \pi L_m^2, \tag{13}$$

where $C_D$ is a dimensionless drag coefficient that quantifies how effectively the artificial magnetosphere (created by the coils and plasma) deflects the incoming solar wind, $u_{sw}$ is the solar wind speed. The efficiency demonstrated in laboratory experiments ($\sim 250$ mN/kW), exceeding ion thrusters.

In Ref. [111] was introduced the *plasma magnet sail* concept, where rotating magnetic fields (RMF) drive azimuthal currents in the plasma, maintaining a large-scale magnetic bubble. The plasma magnet expands with decreasing solar wind pressure, acting like a "magnetic balloon." The Lorentz self-force on plasma currents causes outward expansion until balanced by solar wind pressure. Thrust scales with the effective cross-sectional area, while required input power is reduced due to efficient RMF circuits. In Ref. [112], Quan et al. present a novel study on rotational magnetic sails, a variant of the classic magsail concept. By spinning the magnetic coil, the required sail size can be significantly reduced while maintaining comparable thrust levels in both solar wind and low Earth orbit environments. Using three-dimensional particle tracing simulations, the authors demonstrate that the thrust increases with rotational speed and reaches optimal performance when the magnetic moment of the sail is perpendicular to the incoming particle flow. This study provides insights into enhanced thrust efficiency and attitude stability for future propellantless interplanetary and near-Earth propulsion applications.

The Plasma Magneto-Shell (PMS) is a magnetic sail–like concept developed primarily for aerobraking, atmospheric entry, and deceleration of spacecraft rather than deep-space propulsion. It works by generating a magnetic field around the spacecraft that traps injected plasma, creating an extended "magneto-shell." This inflated plasma structure interacts with the upper atmosphere (or incoming flow), producing drag that allows the spacecraft to decelerate without the need for massive heat shields or large propellant reserves (see [113] and references herein)

TABLE IV: Advantages and disadvantages of different magnetic sail concepts

| Sail type | Advantages | Disadvantages/Challenges |
|---|---|---|
| Classical Magnetic Sail (*MagSail*) | Propellantless propulsion via direct plasma interaction; simple principle using a large superconducting loop; continuous acceleration in solar and interstellar plasma. | Requires very large superconducting loops (tens–hundreds of km); extremely low thrust; deployment and stability of large structures are difficult. |
| Mini-Magnetospheric Plasma Propulsion (*M2P2*) | Plasma injection inflates a magnetic bubble, reducing loop size; smaller structures than classical magsail. | Requires finite onboard propellant for plasma injection; plasma confinement and stability are difficult; performance lower than early MHD predictions. |
| Magnetoplasma Sail (*MPS*) | Plasma injection enhances solar wind interaction; reduced loop size compared to classical design. | Needs continuous plasma supply (propellant); turbulence and diffusion reduce efficiency; still at low Technology Readiness Level (TRL). |
| Plasma Magneto-Shell (*PMS*) | Enables controlled aerobraking and planetary entry; allows propellantless braking against planetary atmospheres. | Not a deep-space propulsion method; requires high power to sustain plasma shell; limited experimental validation. |
| Rotating Magnetic Field (*RMF*) Sail | Generates a magnetic bubble using rotating magnetic fields instead of large coils; reduces reliance on massive superconducting structures; potentially higher thrust-to-mass ratio than classical magsail. | Needs complex rotating magnetic field generators; power-intensive with uncertain efficiency; very low TRL and little experimental validation. |

In Table IV, the advantages and disadvantages of different magnetic sail concepts are presented. The evolution of the working principles of solar magnetic sailing—from the *MagSail* concept to magnetospheric plasma propulsion and the magneto-plasma sail—is reviewed and discussed in Ref. [114], with particular attention to their performance and potential for interplanetary travel.

## VI. ELECTRIC SAILS

The electric sail, known as E-sail, is a novel propellantless propulsion concept that exploits the interaction between charged tethers and the natural solar wind plasma to produce thrust. The E-sail, like the magnetic sail, achieves propellant-free propulsion by deflecting incident space plasma, but uses an electric field instead of a magnetic one. Unlike solar sails, which rely on the momentum transfer of photons, the E-sail utilizes long, positively charged, thin conducting wires (tethers) extended radially from a spinning spacecraft. These tethers are charged to a high positive potential, typically 10–20 kV, using an onboard electron gun, creating an electric field that deflects incident solar wind protons and positively charged ions. The momentum conservation principle ensures that the deflected ions impart momentum to the tethers, producing a continuous thrust on the spacecraft without the consumption of propellant. The original concept was proposed in 2004 by Pekka Janhunen [13] and later following development in [14].

The thrust of an E-sail decreases approximately as $1/r$ with heliocentric distance $r$, rather than $1/r^2$ as in the case of solar photon sails. The physical reasoning is as follows: The solar wind proton number density scales approximately as $n(r) \propto \frac{1}{r^2}$, due to radial expansion. The solar wind speed $v_{\rm sw}$ is approximately constant with $r$, so the dynamic pressure is

$$P_{\rm dyn} \sim n m_p v_{\rm sw}^2 \propto \frac{1}{r^2}. \tag{14}$$

However, the E-sail tether interacts with the solar wind through an electrostatic sheath. The Debye length, which is a characteristic distance in plasmas and charged tether that describes how far a charge's electrostatic influence extends before being shielded by the surrounding collective charges, is given by $\lambda_D = \sqrt{\frac{\varepsilon_0 k_B T}{n_{\rm sw} e^2}} \propto \frac{1}{\sqrt{n_{\rm sw}}} \propto r$. Thus, the effective sheath radius scales as $r_{\rm eff} \propto \sqrt{V}\, r$, where $V$ is the tether potential. Let's consider one tether The thrust of one tether of length $L$ is approximately

$$F \sim P_{\rm dyn} \times (L\, r_{\rm eff}) \sim \left(n_{\rm sw} m_p v_{\rm sw}^2\right) L\, r_{\rm eff}. \tag{15}$$

Substituting the scalings above leads to the thrust force,

$$F \propto \frac{1}{r^2} \times r \ \sim \ \frac{1}{r}. \tag{16}$$

Thus, under standard assumptions (constant $v_{\rm sw}$, constant $T$, and constant tether potential $V$), the thrust of an E-sail decreases approximately as $1/r$.

The E-sail acceleration was studied in different thrust models. According to the thrust model [14], the propulsive acceleration vector $\boldsymbol{a}$ may be expressed as

$$\boldsymbol{a} = a_c \left(\frac{r_\oplus}{r}\right)^{7/6} \hat{\boldsymbol{r}}, \tag{17}$$

where $\hat{\boldsymbol{r}}$ is the Sun–sail unit vector and $a_c$ is the so-called *characteristic acceleration*, i.e., the maximum value of $\|\boldsymbol{a}\|$ at the reference Sun–spacecraft distance 1 AU, $r = r_\oplus$. Proposed an extension of Eq. (17) we can find in Refs. [115–123].

Table V summarizes the advantages and disadvantages (pros and cons) of the E-sail. Reference [124] presents a comprehensive review of the electric sail, covering its concept, deployment strategies, manufacturing methods, structural and thermo-structural analyses, thrust and torque modeling, control strategies, and mission applications. The E-sail's main advantages are its light weight and ability to steer the thrust vector, making it a competitive option alongside conventional thrusters and other propellantless propulsion systems, despite its lower technological maturity. Finally, it should be emphasized that both magnetic and electric sails require an onboard power source: the magnetic sail to sustain the current that generates the magnetic field, and the electric sail to maintain the high potential of the tethers needed to support the electrostatic sheath.

Table VI presents a rough estimation of the accelerations of solar, magnetic, and electric sails, which arise from the electromagnetic and corpuscular components of solar radiation at 1 AU.

## VII. PHYSICAL PRINCIPLES FOR QUANTUM-BASED PROPULSION

In outer space, the environment is characterized by an extremely deep vacuum, almost entirely devoid of matter. A natural question arises: can the phenomenon of quantum vacuum fluctuations, which persist even in empty space, be harnessed in some way for space exploration or propulsion? The Casimir effect, first predicted in 1948 [23, 24], arises when two neutral conducting metal surfaces are brought very close together in a vacuum. It originates from quantum vacuum fluctuations and creates an attractive force between two uncharged, parallel, perfectly conducting plates held at nanometer separations.

TABLE V: Advantages and disadvantages of electric sail

| Advantages | Disadvantages |
|---|---|
| Propellantless propulsion; no onboard fuel required. | Thrust depends on solar wind, which is variable and unpredictable. |
| Capable of continuous long-duration thrust for interplanetary missions. | Deployment of long, lightweight tethers is technically challenging. |
| Thrust can be scaled by increasing tether number or length. | Tether dynamics can induce coning or oscillations, requiring sophisticated control. |
| Low operational costs due to no fuel consumption. | Limited maneuverability near planets due to weak, directional thrust. |
| Rapid response to solar wind by adjusting tether voltage or orientation. | High-voltage tethers require onboard power systems. |
| Potential for fast interplanetary transfers over long distances. | Vulnerable to micrometeoroids and plasma interactions, limiting lifespan. |

TABLE VI: Comparison of approximate order of acceleration of different sail concepts at 1 AU.

| Sail Type | Acceleration (1 AU) | Force Source | Remarks |
|---|---|---|---|
| Solar Sail | 0.1–1 mm/s$^2$ | Solar photon pressure | Depends on reflectivity and mass-to-area ratio. |
| Electric Sail | 0.5–5 mm/s$^2$ | Electrostatic repulsion from solar wind protons | Provides the highest acceleration among sail types; effectiveness decreases with $1/r^2$. |
| Magnetic Sail | 1–10 $\mu$m/s$^2$ | Magnetic deflection of solar wind protons, electrons, ions | Weak acceleration but can operate continuously over large distances. |

The effect is an inherently quantum mechanical phenomenon, stemming from vacuum fluctuations rather than external forces or fields. Because the plates impose boundary conditions on the electromagnetic field, fewer fluctuation modes are allowed between them compared to the outside. This results in a lower vacuum energy density inside the gap, producing a net inward pressure that pushes the plates together, as illustrated in Fig. 2. The Casimir force is extremely small and only becomes significant at sub-micron distances.

The underlying physics is as follows. According to quantum field theory, the vacuum is not truly empty but filled with fluctuating fields that generate virtual particles and zero-point energy. When two perfectly conducting plates are introduced, they restrict the allowed wavelengths of these fluctuations. Since fewer modes exist inside than outside, an imbalance of energy is created, with the higher energy density outside exerting pressure that drives the plates together.

In 1969, Moore in his PhD thesis, part of which was published in [25] predicted *dynamic Casimir effect*. This effect arises when time-dependent boundary conditions, such as rapidly oscillating mirrors or cavities with modulated dielectric properties, convert vacuum fluctuations into real photons. In such systems, virtual photons of the quantum vacuum are promoted to observable radiation due to the nonadiabatic change of the boundary conditions as illustrated in Fig. 2(c). From the propulsion perspective, this effect provides a physical principle for generating thrust: the emitted photons carry momentum, and if their emission is engineered to be asymmetric, a net recoil force on the cavity or structure can result. Although experimentally demonstrated in superconducting microwave circuits, the achievable photon emission rates, and hence the corresponding recoil forces, are currently extremely small. A digest of the main achievements in the wide area of the dynamical Casimir effect, with emphasis on results obtained after 2010, is presented in Ref. [125], while Refs. [126, 127] provide review for earlier studies.

If the distance between the boundaries $L(t)$ slowly varies with time (in the sense that $dL/dt << c$), a generalize

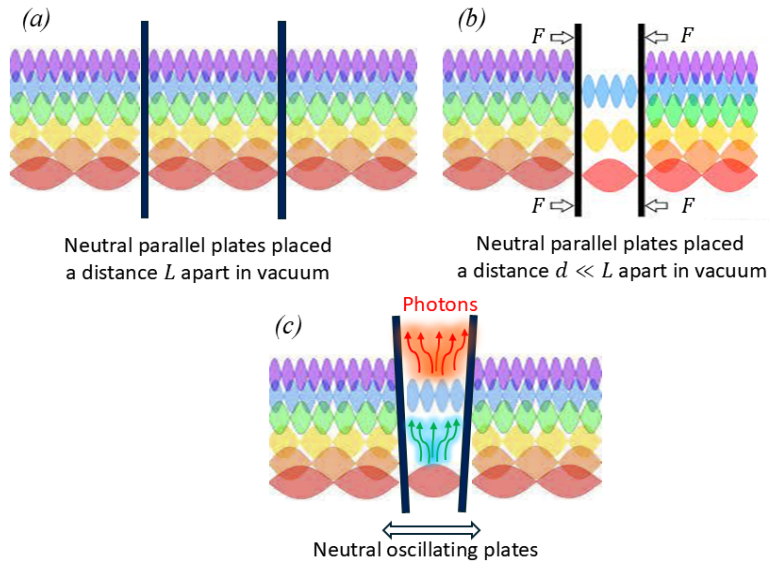

FIG. 2: Schematic illustration of the Casimir effect using the quantum electromagnetic model. ($a$) At macroscopic plate separations $L$, quantum vacuum fluctuations have symmetric wavelengths and no net force arises. ($b$) At sub-micron distances $d$ between the plates, quantum vacuum fluctuations are restricted, leading to an attractive Casimir force $F$ that pushes the plates together due to the higher energy density of modes outside. ($c$) The dynamic Casimir effect occurs when boundary conditions change in time, converting vacuum fluctuations into real photons. The emitted photons carry momentum, and if emission is asymmetric, a net recoil force can be produced. Distances between plates are not to scale.

formula for the thrust force reads as

$$F = -\frac{\pi \hbar c}{24 L^2(t)} \left[ 1 + \left(\frac{\dot{L}}{c}\right)^2 \left(\frac{7}{3} - \frac{1}{\pi^2}\right) - \frac{L\ddot{L}}{c^2}\left(\frac{2}{3} - \frac{2}{\pi^2}\right) \right]. \tag{18}$$

The Casimir force in a cavity moving with a constant acceleration was calculated in [129].

The paper [130] explores whether the vacuum—predicted by quantum electrodynamics to contain fluctuating virtual photons—can be harnessed for propellantless spacecraft propulsion. The author outlines a thought experiment spacecraft powered by vibrating mirrors interacting with the quantum vacuum. In principle, this could produce thrust through asymmetric photon emission, although the resulting accelerations are extremely small with current technology. Nevertheless, the dynamic Casimir effect illustrates how quantum vacuum fluctuations can, in principle, be harnessed for propellantless propulsion concepts [130, 131].

Thus, the Casimir effect, arising from quantum vacuum fluctuations, has been proposed as a potential mechanism for propellantless propulsion. Approaches include: (i) *lateral Casimir forces* induced by engineered surface inhomogeneities or gratings, which can generate tiny lateral forces on nearby nano-objects [128]; (ii) the *dynamic Casimir effect*, where modulated cavities convert vacuum fluctuations into real photons, yielding recoil thrust [131]; and (iii) *symmetry-breaking effects* using rotating particles or magneto-electric materials to bias vacuum fluctuations directionally [132, 133]. While these effects are real at nanoscales, their magnitude is currently far too small for macroscopic spacecraft propulsion [134, 135]. Nonetheless, they provide valuable insight into quantum vacuum interactions and could enable nano-/micro-scale actuation and manipulation.

Anyway, new hypothetical propulsion concepts have been proposed that seek to leverage the Casimir effect to generate propellantless thrust by exploiting the vacuum energy of space, rather than relying on conventional engines, although such ideas remain highly speculative and far from practical realization.

Does a charged high-voltage capacitor produce thrust? If true, this could open the way to a novel space propulsion scheme, which would be of considerable interest [136]. Several claims have appeared in the literature suggesting that a charged high-voltage capacitor can generate thrust. More recently, theoretical models have been proposed to support such an electrostatic propulsion scheme [137–140]. The authors of Ref. [141] reviewed these models, which predict that electrostatic effects could cause the self-acceleration of a dielectric when polarized, and described an experimental setup designed to measure possible weight changes or forces in capacitors charged up to 10 kV, while carefully eliminating side effects associated with high voltages. In Fig. 3, the main capacitor configurations investigated are shown. In these models, the electrostatic field is predicted to generate a gravitational field that accelerates the dielectric. However, no force was detected across a variety of configurations, effectively ruling out most proposed theories by many orders of magnitude.

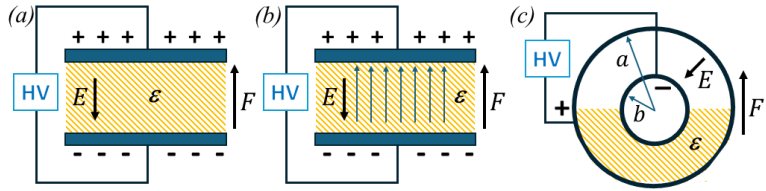

FIG. 3: Illustration of the force $F$ created by the electric field $E$ that acts on the different capacitor configurations with dielectric permittivity $\epsilon$. ($a$) A simple parallel-plate capacitor. ($b$) A parallel-plate capacitor with leaking current. ($c$) A spherical capacitor with radii $a$ and $b$ half-filled with a dielectric. No force $F$ was detected for these configurations, ruling out most theories by many orders of magnitude [141].

## VIII. CONCLUDING REMARKS

Chemical rockets remain the dominant technology for space exploration, but they suffer from fundamental limitations in efficiency and achievable velocity. This review explores astrophysical alternatives—such as gravity-assist maneuvers and solar, magnetic, and electric sails—that eliminate the need for onboard propellant, as well as quantum field–based approaches that aim to extract energy from the vacuum space.

The gravity-assist manoeuvre has fundamentally transformed interplanetary space exploration by allowing spacecraft to achieve significant increases in heliocentric velocity without additional propellant. Its theoretical development progressed from early conceptual ideas to quantitative formalism and eventually to widespread application in mission design. By transferring a small fraction of a planet's orbital momentum to the spacecraft, gravity-assist flybys efficiently reduce flight times and fuel requirements, enabling successful missions to the outer planets and beyond, including Voyager, Galileo, Cassini, New Horizons, Solar Orbiter, and Parker Solar Probe.

Solar sailing is a mature and elegant concept that leverages photon momentum transfer to enable propellant-free propulsion. Its foundation lies in well-established electromagnetic theory, confirmed experimentally since the early 20th century. Demonstrations such as *IKAROS* and *LightSail-2* have proven its feasibility, while ongoing research continues to expand applications to non-Keplerian orbits, deep-space exploration, and even fundamental physics tests. Despite challenges related to sail deployment, material durability, and low thrust levels, solar sails remain a promising technology for long-duration interplanetary and interstellar missions. Thus, solar and beam-driven sails provide propellant-free propulsion, enabling continuous acceleration, unique orbital configurations, and interplanetary or interstellar missions that are unattainable with conventional chemical rockets.

Magnetic sails represent a versatile class of propellantless spacecraft propulsion systems that leverage the interaction between artificial magnetic fields and ambient plasma to generate thrust. From the original MagSail concept to advanced variants such as M2P2, MPS, RMF sails, and plasma magneto-shells, each design offers trade-offs between structural complexity, thrust efficiency, and operational environment. While classical designs achieve thrust via large superconducting loops, plasma-assisted and rotational field concepts aim to reduce structural demands and improve performance. Despite technical challenges, magnetic sails hold significant promise for continuous, propellant-free acceleration in both interplanetary and near-Earth applications, with potential extensions to aerobraking and orbital maneuvering.

Electric sails offer a propellantless propulsion method by using long, charged tethers to deflect solar wind ions, producing continuous thrust without onboard fuel. Their thrust scales approximately as $1/r$ with distance from the Sun, enabling efficient long-duration interplanetary travel. While deployment of long tethers and dependence on variable solar wind pose technical challenges, E-sails provide a lightweight, low-cost alternative to conventional propulsion and remain a promising technology for fast, propellant-free space missions.

Hypothetical propulsion concepts, sometimes very speculative, have been proposed that attempt to leverage the Casimir effect to generate propellantless thrust by exploiting the vacuum energy of space, rather than relying on a conventional engine. In particular, the dynamic Casimir effect provides a possible route for such concepts: when boundary conditions, such as the position of mirrors or dielectric properties of cavities, change rapidly in time, virtual photons from the quantum vacuum can be converted into real photons. Since these photons carry momentum, their asymmetric emission could, in principle, impart a net recoil force on a cavity or spacecraft structure. Although current experimental realizations demonstrate only extremely small photon emission rates, the principle illustrates a potential pathway toward propellantless propulsion systems based on harnessing quantum vacuum fluctuations.

Hopefully, studies on vacuum propulsion during the next 100 years will find amusing our failure to perceive the key issues and see clearly how quantum propulsion should be done.

---